\documentclass[12pt]{iopart}

\usepackage{graphicx}

\usepackage{dcolumn,multirow,color}

\begin{document}

\title[Communication cliques in calling networks]{Communication cliques in mobile phone calling networks}

\author{Ming-Xia Li$^{1,2,3}$, Wen-Jie Xie$^{2,3}$, Zhi-Qiang Jiang$^{2,3}$ and Wei-Xing Zhou$^{1,2,3}$}
\address{$^1$ Department of Mathematics, East China University of Science and Technology, Shanghai 200237, China}
\address{$^2$ School of Business, East China University of Science and Technology, Shanghai 200237, China}
\address{$^3$ Research Center for Econophysics, East China University of Science and Technology, Shanghai 200237, China}
\ead{zqjiang@ecust.edu.cn; wxzhou@ecust.edu.cn}

\begin{abstract}
  People in modern societies form different social networks through numerous means of communication. These communication networks reflect different aspects of human's societal structure. The billing records of calls among mobile phone users enable us to construct a directed calling network (DCN) and its Bonferroni network (SVDCN) in which the preferential communications are statistically validated. Here we perform a comparative investigation of the cliques of the original DCN and its SVDCN constructed from the calling records of more than nine million individuals in Shanghai over a period of 110 days. We find that the statistical properties of the cliques of the two calling networks are qualitatively similar and the clique members in the DCN and the SVDCN exhibit idiosyncratic behaviors quantitatively. Members in large cliques are found to be spatially close to each other. Based on the clique degree profile of each mobile phone user, the most active users in the two calling networks can be classified in to several groups. The users in different groups are found to have different calling behaviors. Our study unveils interesting communication behaviors among mobile phone users that are densely connected to each other.
%
\end{abstract}



\maketitle

\section{Introduction}
\label{sec:introduction}

Human in modern societies are connected through numerous ties to form complex networks which have fine localized structures \cite{Albert-Barabasi-2002-RMP,Newman-2003-SIAMR,Boccaletti-Latora-Moreno-Chavez-Hwang-2006-PR,Fortunato-2010-PR,Estrada-Hatanoe-Benzi-2012-PR,Holme-Saramaki-2012-PR}. The definition and identification of ties between individuals are not obvious. Taking friendship as an example, the conventional way is self-reporting through questionnaire survey with the main shortcoming of small sample \cite{Eagle-Penland-Lazer-2009-PNAS}. Even if we are able to define and record a measure of closeness among individuals according to their mutual communication and social activities, we still need to further preset a closeness threshold for the definition of friendship \cite{Xie-Li-Jiang-Zhou-2014-SR}. In recent years, the availability of mobile phone billing data provides us a new proxy to investigate human social ties through their communications. The inference of friendship network structure by using mobile phone data offers a promising alternative method to self-reporting survey \cite{Eagle-Penland-Lazer-2009-PNAS}.

Mobile phone data contain unevenly sampled temporal and spatial information of phone users. Hence, the large-scale mobility patterns of human populations can be investigated and predicted in very high accuracy \cite{Gonzalez-Hidalgo-Barabasi-2008-Nature,Kang-Ma-Tong-Liu-2012-PA,Song-Qu-Blumm-Barabasi-2010-Science,Lu-Bengtsson-Holme-2012-PNAS,Simini-Gonzalez-Maritan-Barabasi-2012-Nature}. In addition, the temporal patterns of human's communication dynamics are also unveiled to exhibit bursty behaviors with non-Poissonian characteristics \cite{Gonzalez-Hidalgo-Barabasi-2008-Nature,Hong-Han-Zhou-Wang-2009-CPL,Candia-Gonzalez-Wang-Schoenharl-Madey-Barabasi-2008-JPAMT,Wu-Zhou-Xiao-Kurths-Schellnhuber-2010-PNAS,Zhao-Xia-Shang-Zhou-2011-CPL,Karsai-Kaski-Kertesz-2012-PLoS1,Karsai-Kaski-Barabasi-Kertesz-2012-SR,Jiang-Xie-Li-Podobnik-Zhou-Stanley-2013-PNAS,Kovanen-Kaski-Kertesz-Saramaki-2013-PNAS,Quadri-Zignani-Capra-Gaito-Rossi-2014-PLoS1}. These studies provide rich information for improving urban planning, traffic design, crowd evacuation, tourism recommendation, and so on.

Very rich structural properties of mobile phone calling networks have been uncovered \cite{Onnela-Saramaki-Hyvonen-Szabo-deMenezes-Kaski-Barabasi-Kertesz-2007-NJP,Li-Jiang-Xie-Micciche-Tumminello-Zhou-Mantegna-2014-SR}. Palla et al. found that large-size groups persist longer if the proportion of migrated members is large and small-size groups are more stable if their members are less immigrant \cite{Palla-Barabasi-Vicsek-2007-Nature}. Onnela et al. confirmed the importance of weak ties in social networks \cite{Onnela-Saramaki-Hyvonen-Szabo-Lazer-Kaski-Kertesz-Barabasi-2007-PNAS}. It is also found that the European calling network exhibits communities \cite{Kumpula-Onnela-Saramaki-Kaski-Kertesz-2007-PRL,Jo-Pan-Kaski-2011-PLoS1}. On the micro level, triadic motifs are overrepresented or underrepresented in calling networks \cite{Li-Palchykov-Jiang-Kaski-Kertesz-Micciche-Tumminello-Zhou-Mantegna-2014-NJP}, and the investigation of temporal motifs unveils the presence of temporal homophily in human communications \cite{Kovanen-Kaski-Kertesz-Saramaki-2013-PNAS}.

An interesting local network structure is clique in which all nodes are connected completely with each other. The properties of cliques contain specific behaviors of certain mobile phone users. Onnela et al. found that high-order cliques exist in the European calling network and members in cliques have heavier interactions as revealed by clique intensity and coherence \cite{Onnela-Saramaki-Hyvonen-Szabo-deMenezes-Kaski-Barabasi-Kertesz-2007-NJP}. To be self-consistent, we describe briefly some key concepts in graph theory that are investigated in this work. A directed graph is called complete if every two distinct nodes $i$ and $j$ are connected by two directed links $i\to{j}$ and $j\to{i}$. A simplex of a graph is any complete subgraph \cite{Prisner-1995}. We first identify all simplices as follows. We find out all simplices of size 3 by adopting a widely used method proposed in \cite{Milo-ShenOrr-Itzkovitz-Kashtan-Chklovskii-Alon-2002-Science}. For each triadic simplex, we investigate each of the remaining nodes to check if it is bidirectionally connected to every nodes of the simplex. In this way, we identify all simplices of size 4. This procedure continues until that no simplex of size $k+1$ is identified from simplices of size $k$. Cliques are inclusion-maximal simplices of a graph \cite{Prisner-1995}. We denote $C_k$ the $k$-cliques of size $k$. The set of cliques is determined by removing all simplices of size $k$ that are subgraphs of other simplices of size $k+1$. For instance, if a triadic simplex is a subgraph of a quaternary simplex, then it is not a 3-clique.

In this work, we will perform a detailed analysis on the statistical properties of cliques in a directed calling network of Chinese mobile phone users and its Bonferroni version in which the links are filtered by a statistical validation test \cite{Tumminello-Micciche-Lillo-Piilo-Mantegna-2011-PLoS1}. The rest of this paper is organized as follows. Section \ref{S1:Data} describes the data set used in this work. Section \ref{S1:Topological} presents topological properties of cliques and Section \ref{S1:Behavior} studies behavioral properties of mobile phone users from the view angle of cliques. Section \ref{S1:Discussion} discusses and summarized the results.

\section{Data description}
\label{S1:Data}

The data set was provided by a Chinese mobile phone operator for scientific research. The phone numbers were anonymized and de-identified prior to analysis in the sense that these numbers were one-to-one mapped into natural numbers. The data set analyzed in this paper contains the call records of more than nine million phone numbers from one of the three mobile operators in China. It covers two separated periods: one is from June 28, 2010 to July 24, 2010 and the other is from October 1, 2010 to December 31, 2010. Due to unknown reasons, the records of a few hours are missing on eight days (October 12, November 6, 21, 27, and December 6, 8, 21, 22). These days are thus excluded from our investigation sample.

The main information used to construct calling networks is about who calls whom. A directed calling network (DCN) is constructed \cite{Li-Jiang-Xie-Micciche-Tumminello-Zhou-Mantegna-2014-SR}, where the nodes are the mobile phone users (more precisely, mobile phone numbers) and the links are drawn from call initiators (callers) to call receivers (callees). The DCN has 4,032,884 nodes and 16,753,635 directed links. We then perform statistical tests on the links based on a method proposed in Ref.~\cite{Tumminello-Micciche-Lillo-Piilo-Mantegna-2011-PLoS1}, which has been applied to a number of complex networks in diverse fields \cite{Tumminello-Micciche-Lillo-Piilo-Mantegna-2011-PLoS1,Tumminello-Micciche-Lillo-Varho-Piilo-Mantegna-2011-JSM,Tumminello-Lillo-Piilo-Mantegna-2012-NJP,Curme-Tumminello-Mantegna-Stanley-Kenett-2014-XXX,Hatzopoulos-Iori-Mantegna-Micciche-Tumminello-2014-QF,Li-Palchykov-Jiang-Kaski-Kertesz-Micciche-Tumminello-Zhou-Mantegna-2014-NJP}. The links of the DCN that do not pass the test are removed, which results in a statistically validated directed calling network (SVDCN) known as Bonferroni network. A detailed description of the construction of the DCN and SVDCN can be found in Ref.~\cite{Li-Jiang-Xie-Micciche-Tumminello-Zhou-Mantegna-2014-SR}. The resulting SVDCN has 2,410,757 nodes and 2,453,678 edges.

\section{Topological properties}
\label{S1:Topological}

\subsection{Clique size}

\begin{figure}[htp]
  \centering
  \includegraphics[width=0.99\textwidth]{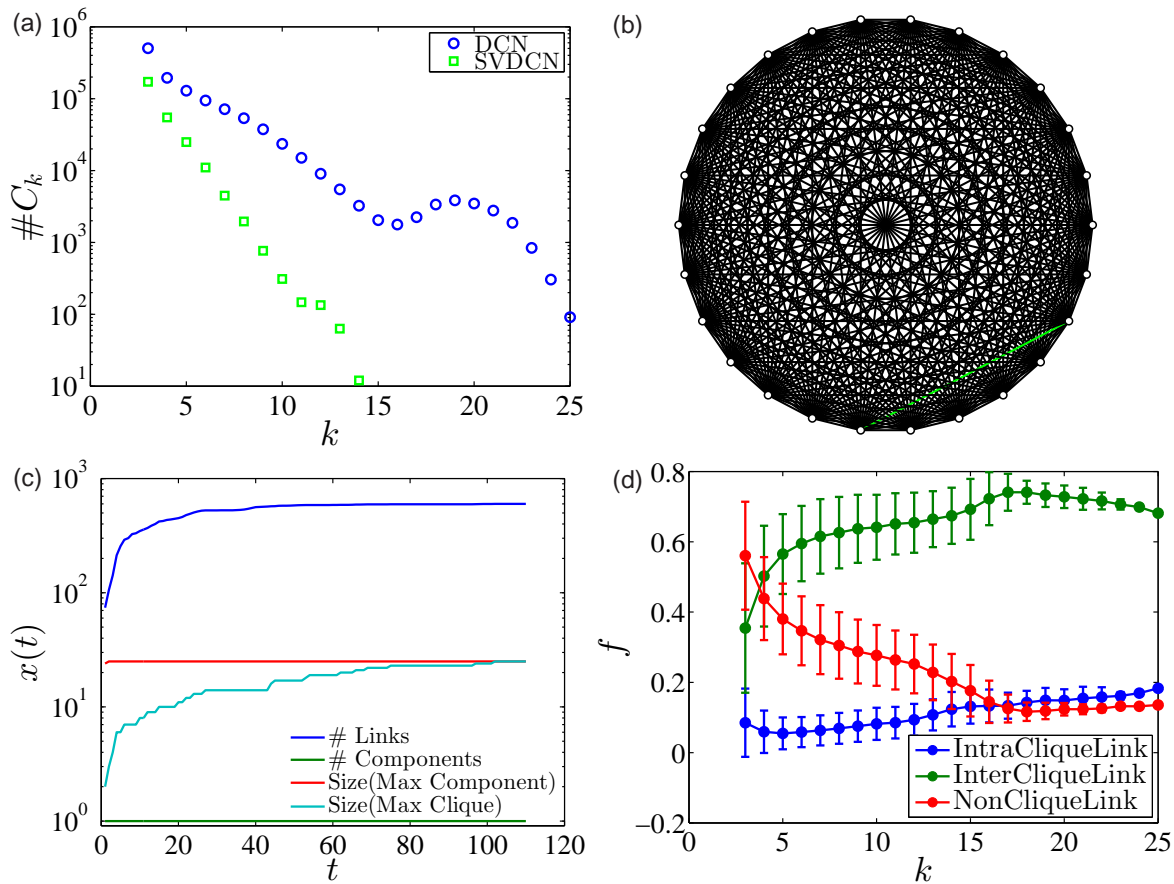}
  \caption{{\textbf{Clique size.}} (a) Number of $k$-cliques, $\#C_k$, as a function of $k$ for the DCN and the SVDCN. (b) An example of two 25-cliques with significant overlaps. After adding a directed link (the green line), it becomes a $26$-clique. (c) Evolution of a 25-clique. (d) Fraction of three types of links connected to the nodes in different cliques.}
  \label{Fig:CliqueSize}
\end{figure}

Figure \ref{Fig:CliqueSize}(a) illustrates the number of $k$-cliques, $\#C_k$, as a function of clique size $k$ for the DCN and SVDCN. For the DCN, the size of the largest cliques is $k=25$. When $k\leq16$, $\#C_k$ decreases with $k$. Surprisingly, there is a hump for large $k$ values. The cliques of the same size may have large overlaps, as shown in Fig.~\ref{Fig:CliqueSize}(b), in which two $25$-cliques will form a $26$-clique if one link is added. For the SVDCN, the size of the largest cliques is $k=14$ and there is also a milder hump at the right tail. The clique size profile of the DCN has an exponential form when $k<12$. Comparing these two profiles, we find that a major proportion of the links in the DCN cliques are not statistically significant.

To understand the evolution of the largest DCN cliques, we investigate the temporal evolution of the directed calling network of these 25 nodes for all the 25-cliques on a daily base. The number of links, the number of components, the size (or the number of nodes) of the largest component, and the size (or the number of nodes) of the largest clique are determined for each evolving directed calling network on each day. These quantities evolve in similar ways for all the 25-cliques and the results for a randomly chosen 25-clique are shown in Fig. \ref{Fig:CliqueSize}(c). We find that these 25 nodes are connected since the first day such that the number of component is 1 and the size of the largest component is 25. The number of links and the size of the largest clique increases to maturity.

For the nodes of a $k$-clique in the DCN, the links (at least one node of a link belonging to this $k$-clique) can be divided into three types: intra-clique link whose two nodes belonging to the same $k$-clique, inter-clique link with one node belonging to another clique of any size, and non-clique link whose second node does not belong to any clique. We determine the fractions of the three types of link for each $k$-clique. The means of the three fractions, $f_1(k)$, $f_2(k)$, and $f_3(k)$, are calculated over all $k$-cliques, as shown in Fig. \ref{Fig:CliqueSize}(d). The fractions of intra-clique and inter-clique links increase with $k$, while the fraction of non-clique links decreases. This observation suggests that cliques with large size have high overlapping level and cliques with small size are separated. It means that users in large cliques are more likely to interact with each other even though they are in different cliques.

\subsection{Clique intensity}

Combining the network structure and the interaction strength among mobile phone users, we can determine the the subgraph intensity. The intensity of subgraph $C_k$ is given by the geometric mean of its weights:
\begin{equation}
  i(k) = \left(\prod_{(i,j) \in C_k} w_{ij}\right)^{1/l_k}.
  \label{Eq:Intensity}
\end{equation}
where $l_k=k(k+1)$ is the number of links in $C_k$ and $w_{ij}$ is the total duration of calls from user $i$ to user $j$. The unit of intensity is the same as the unit of network weights. Clique intensity is a measure of average call duration among the users in the same cliques.

\begin{figure}[htp]
  \centering
  \includegraphics[width=0.99\textwidth]{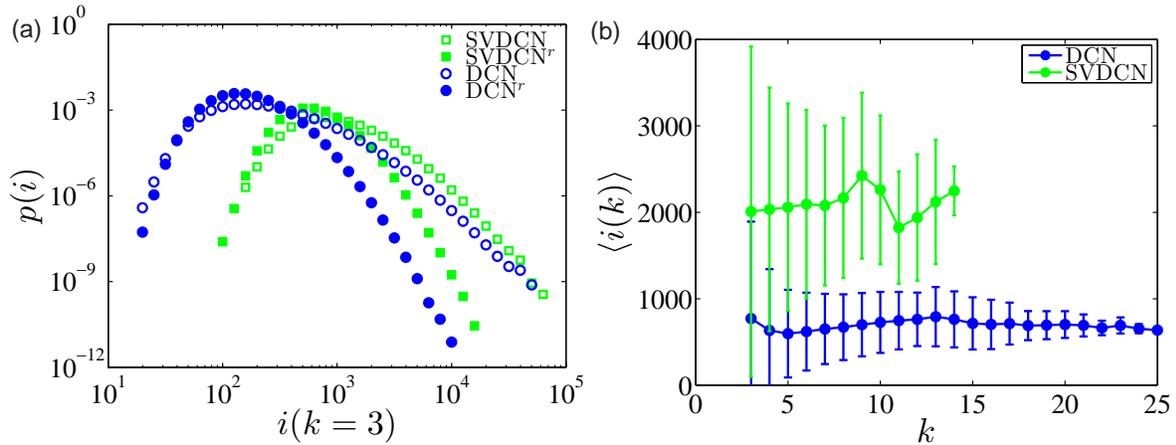}
  \caption{{\textbf{Clique intensity.}} (a) Probability distribution of 3-clique intensity for the DCN and the SVDCN and their weight-randomized networks DCN$^r$ and SVDCN$^r$. The random results are obtained by averaging intensity of 100 weight shuffled networks while keeping the network topology. (b) Dependence of the average intensity for $k$-cliques as a function of clique size $k$.}
  \label{Fig:Intensity}
\end{figure}

Figure \ref{Fig:Intensity}(a) plots the empirical intensity probability distributions of the 3-cliques in the DCN and the SVDCN. The 3-cliques of the SVDCN have higher intensity than those of the DCN. We further shuffle the weights in the networks to obtain reference results, which removes weight correlations while keeps the network topology. The shuffled results are also shown in Fig. \ref{Fig:Intensity}(a). The distribution curves for the shuffled results are narrower than the original ones. Especially, there are much less 3-cliques with high intensity for the shuffled networks, indicating that the communication intensity of real people in a clique can be much higher than by chance. The results for other cliques with high $k$ values are similar. These observations are qualitatively the same as the European calling network, where the authors investigated the original calling network without filtering out insignificant links \cite{Onnela-Saramaki-Hyvonen-Szabo-deMenezes-Kaski-Barabasi-Kertesz-2007-NJP}.

Figure \ref{Fig:Intensity}(b) presents the average intensity of cliques with different size $k$ for the DCN and the SVDCN. There is no significant trend in the average intensity in regard of clique size for both networks. However, the cliques have much higher average intensity in the SVDCN than in the DCN. It means that people in the cliques of the statistically validated network have much more interactions among each other than people in the cliques of the original DCN. We also observe that the clique intensity fluctuation decreases with clique size $k$.

\subsection{Clique coherence}

The coherence of clique $C_k$ is defined by the arithmetic mean of the weights as follows:
\begin{equation}
\label{EQ:Coherence}
q(k) = i(k) l_k /\sum_{(i,j) \in C_k} w_{ij},
\end{equation}
where $q(k) \in [0,1]$ and it is close to unity only if the weights of the clique do not differ much from each other. Clique coherence is to quantify the homogeneity of wights in the clique. We also randomize the weights of the DCN and the SVDCN, identify the cliques of the randomized networks, and calculated the associated clique coherence.

\begin{figure}[htp]
  \centering
  \includegraphics[width=0.99\textwidth]{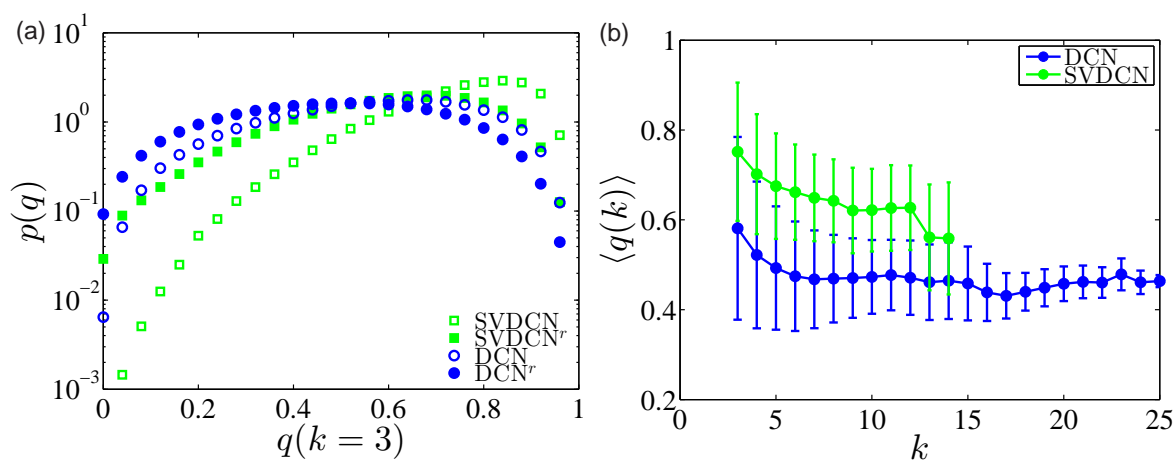}
  \caption{{\textbf{Clique coherence.}} (a) Probability distribution of 3-clique coherence for the DCN and the SVDCN and their weight-randomized networks DCN$^r$ and SVDCN$^r$. The random results are obtained by averaging the coherence of 100 weight randomized networks while keeping the network topology. (b) Dependence of the average coherence for $k$-cliques as a function of clique size $k$. }
  \label{Fig:Coherence}
\end{figure}

Figure \ref{Fig:Coherence}(a) shows the coherence of the 3-cliques in the two calling networks. It is observed that the 3-cliques in the original networks are more coherent than their randomized counterparts because the realistic cliques have larger probabilities for large $q$ values. Moreover, the proportion of high coherence cliques is larger in the SVDCN than in the DCN. These observations are qualitatively the same as the European calling network \cite{Onnela-Saramaki-Hyvonen-Szabo-deMenezes-Kaski-Barabasi-Kertesz-2007-NJP}.

The average coherence of $k$-cliques for the two calling networks is shown in Fig.~\ref{Fig:Coherence}(b). It is clear that the cliques in the SVDCN are more coherent than those in the DCN. This finding suggests that users belonging to the same clique in the SVDCN have closer comparable interaction strengths than their DCN counterparts, which is reminiscent of triadic dependence motifs or collaboration networks in which the difference of social ranks is very low among agents \cite{Bhattacharya-Dugar-2014-MS,Xie-Li-Jiang-Zhou-2014-SR}. There is a mild decreasing trend in the average coherence curve when the clique size is small ($<6$) for the DCN. The decreasing trend in the $q(k)$ is more pronounced for the SVDCN. This decreasing trend is due to the fact that it is hard for users in large cliques to be very coherent. In addition, the standard deviation of the coherence decrease with $k$ for the DCN, which is less clear for the SVDCN.

%

\subsection{Clique's degree and strength}

Given a graph $G(V,E)$ and a $k$-clique $C_k$ of $G$, the edge neighborhood of $C_k$ can be defined as the total number of edges running between $C_k$ and $V\backslash C_k$, being denoted by $N(C_k)$ \cite{Martins-2012-COR}. We call it {\textit{clique's degree}}, which should not be confused with {\textit{node's clique degree}} defined below. For simplicity, to determine $N(C_k)$ of a $k$-clique, we can treat this clique as a node and determine its degrees such as in-degree and out-degree in which the self-loops (intra-clique links) are excluded.

\begin{figure}[htp]
  \centering
  \includegraphics[width=0.99\textwidth]{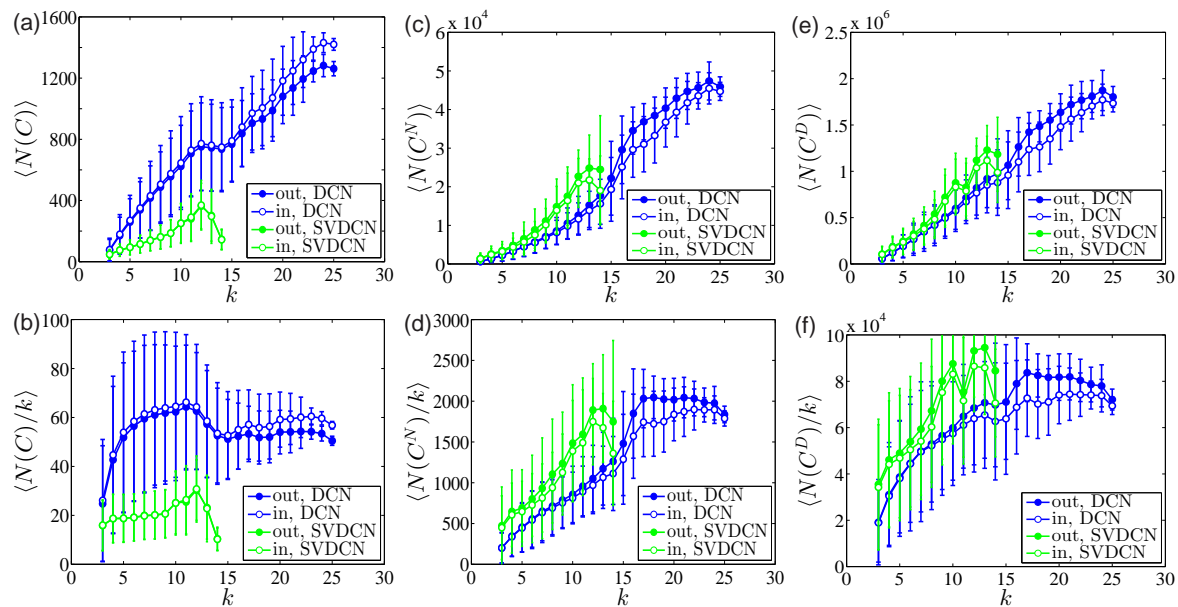}
  \caption{{\textbf{Clique's degrees and strengths for the DCN and the SVDCN.}} (a) Average clique's out-degree and in-degree as a function of clique size. (b) Average clique's out-degree and in-degree per capita as a function of clique size. (c) Average clique' out-strength and in-strength based on call numbers as a function of clique size. (d) Average clique' out-strength and in-strength per capita based on call numbers as a function of clique size. (e) Average clique' out-strength and in-strength based on call durations as a function of clique size. (f) Average clique' out-strength and in-strength per capita based on call durations as a function of clique size. }
  \label{Fig:Clique's:Degree:Strength}
\end{figure}

Figure~\ref{Fig:Clique's:Degree:Strength}(a) shows the average clique's in-degree and out-degree of $k$-cliques as a function of clique size $k$ for the two calling networks. For each calling network, the average clique's in-degree $\langle{N^{\textrm{in}}(C_k)}\rangle$ is equal to or slightly larger than the average clique's out-degree $\langle{N^{\textrm{out}}(C_k)}\rangle$, indicating that the average number of users called the clique members is larger than the number that the clique numbers called non-clique users. In addition, both curves for the DCN have an increasing trend, while the two curves for the SVDCN increase when $k\leq12$ and then decrease. Figure~\ref{Fig:Clique's:Degree:Strength}(b) shows the average clique's in-degree and out-degree per capita. The two curves for the DCN have two humps and slightly differ from each other. A 3-clique member has respectively about 25 non-clique incoming and outgoing contacts, which is significantly smaller than other $k$-clique members. Usually a clique member has 40-60 incoming or outgoing contacts. Clique members in the DCN have more incoming contacts than outgoing contacts and the difference between $\langle{N^{\textrm{in}}(C_k)}/k\rangle$ and $\langle{N^{\textrm{out}}(C_k)}/k\rangle$ increases with clique size $k$. In contrast, the two curves for the SVDCN are unimodal and almost overlap. The average degrees for both networks increase with clique size $k$ and reach the maximum at $k=12$.

Figure \ref{Fig:Clique's:Degree:Strength}(c) shows the average clique's in-strength and out-strength based on call numbers, where $N^{\mathrm{in}}(C^N)$ is defined as the total number of calls from non-clique users to clique members and $N^{\mathrm{out}}(C^N)$ is defined as the total number of calls from clique members to non-clique users. We find that the in-strength is slightly smaller than the out-strength and the four $N(C^N)$ curves have an increasing trend with a turndown for the largest cliques. An interesting feature is that the cliques of the SVDCN have greater in-strength and out-strength than the cliques of the DCN. The average clique's in-strength and out-strength per capita based on call numbers are presented in Fig. \ref{Fig:Clique's:Degree:Strength}(d), which have similar properties as in Fig. \ref{Fig:Clique's:Degree:Strength}(c) except that the two curves for the DCN turn downward earlier. We also defined call duration based strengths $N(C^D)$ and show the results in Fig. \ref{Fig:Clique's:Degree:Strength}(e) and (f). The results of $N(C^D)$ are similar to $N(C^N)$. These observations indicate that mobile phone users in large cliques are usually have heavier communications with others, although their numbers of contacts do not vary much.

\subsection{Distributions of clique degree and simplex degree}

\begin{figure}[htp]
  \centering
  \includegraphics[width=0.99\textwidth]{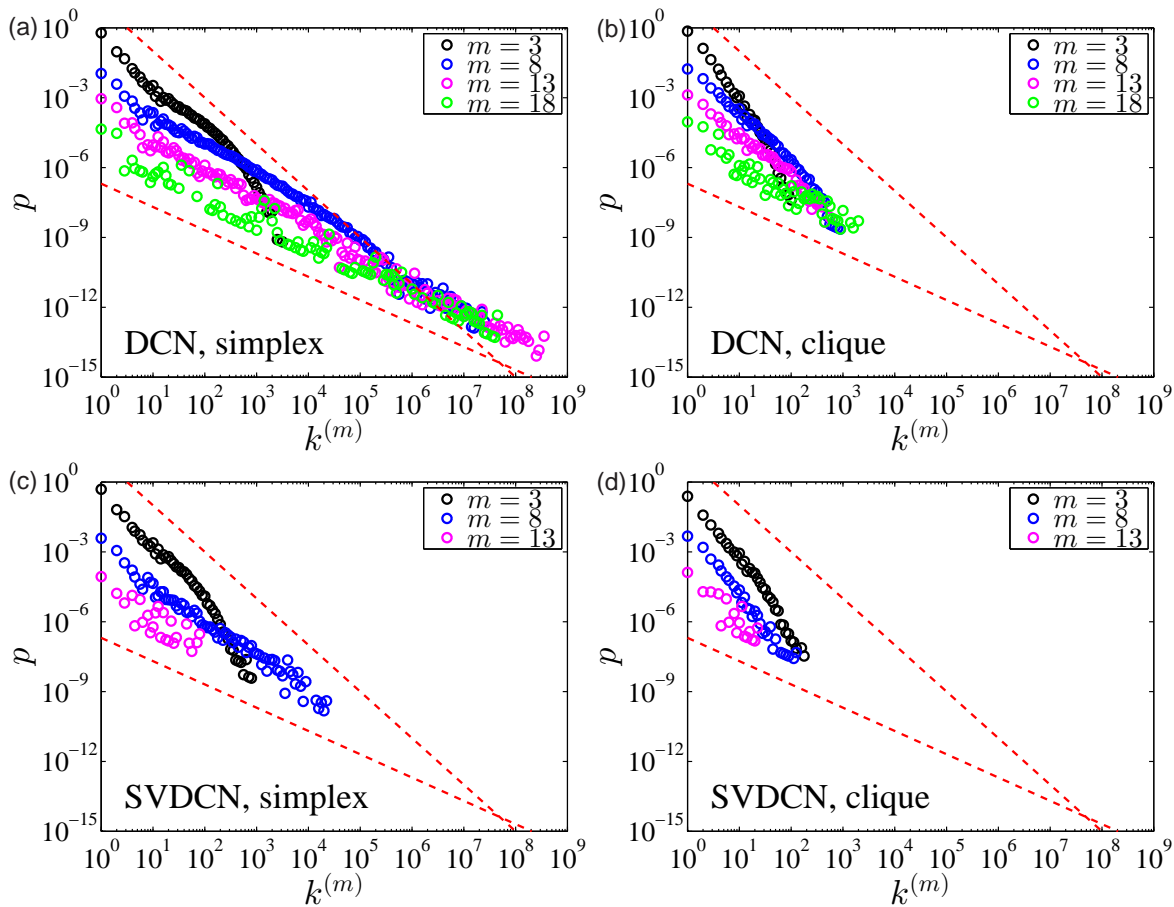}
  \caption{{\textbf{Evolution of distributions of $m$-clique degrees and $m$-simplex degrees for the DCN and the SVDCN.}} (a) Simplices of the DCN. (b) Cliques of the DCN. (c) Simplices of the SVDCN. (d) Cliques of the SVDCN. The dashed lines have a power-law exponent of 2, while the solid lines have a power-law exponent of 1.}
\label{Fig:CliqueDegreeDistribution}
\end{figure}

The $m$-simplex degree $k_i^{(m)}$ of a node $i$ is defined as the number of $m$-simplices containing node $i$ \cite{Lim-Peng-1981-JGT}. Clearly, a 2-simplex is an edge and $k_i^{(2)}$ equals the degree $k_i$ of node $i$. For simplicity, we can call the $m$-simplex degree by simplex degree. The simplex degree distributions of many real networks have been studied and most of them have power-law tails \cite{Xiao-Ren-Qi-Song-Zhu-Yang-Jin-Wang-2007-PRE,Yang-Wang-Liu-Han-Zhou-2008-CPL,Fagiolo-Reyes-Schiavo-2009-PRE,Feng-Fu-Xu-Zhou-Chang-Wang-2012-PA}. We note that, in these studies \cite{Lim-Peng-1981-JGT,Xiao-Ren-Qi-Song-Zhu-Yang-Jin-Wang-2007-PRE,Yang-Wang-Liu-Han-Zhou-2008-CPL,Fagiolo-Reyes-Schiavo-2009-PRE,Feng-Fu-Xu-Zhou-Chang-Wang-2012-PA}, the authors used the term ``clique degree'' instead of ``simplex degree''. The simplex degree distributions of four different simplices for the DCN are illustrated in Figure~\ref{Fig:CliqueDegreeDistribution} (a). These distributions can be split into two parts, a power-law bulk and a humped tail. The power-law exponent is close to 2, while the hump becomes broader with the increase of simplex size and reaches a power law of exponent 1 for the 18-simplices. The three simplex distributions for the SVDCN have similar shapes as those for the DCN.

The $m$-clique degree $k_i^{(m)}$ of a node $i$ can be defined similarly as the number of different $m$-cliques containing $i$. In this case, the 2-clique degree $k_i^{(2)}$ is less than the degree $k_i$ of node $i$. As shown in Fig. \ref{Fig:CliqueDegreeDistribution}, the clique degree distributions are narrower than the corresponding simplex degree distributions and have power-law shapes.

\subsection{Multicliqual edges}

\begin{figure}[htp]
  \centering
  \includegraphics[width=0.99\textwidth]{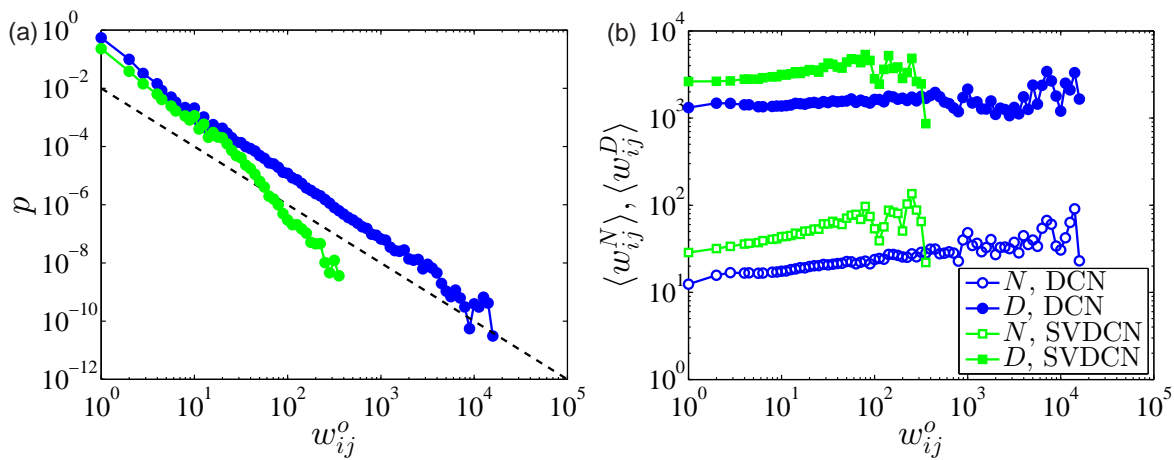}
  \caption{{\textbf{Multicliqual edges in the DCN and the SVDCN.}} (a) Probability distribution for the overlapping level $w_{ij}^o$ of each link. (b) Average link weight as a function of the overlapping level $w_{ij}^o$.
}
\label{Fig:MultiCliqueEdges}
\end{figure}

A multicliqual edge is an edge belongs to more than one clique \cite{Prisner-1995}. We can determine the number of cliques that an edge belongs to as an overlapping level for each link, which is denoted by $w_{ij}^o$. A link $(i,j)$ is more important to the network if its $w_{ij}^o$ value is large. Figure \ref{Fig:MultiCliqueEdges}(a) shows the probability distributions of $w_{ij}^o$ of each link in the two calling networks, including $w_{ij}^o=1$. Both distributions have a power law in the bulk followed by a power-law tail. The power law exponents in the bulk for both calling networks are similar and slightly larger than 2. In contrast, the DCN has a longer tail than the SVDCN which decays slower. Figure \ref{Fig:MultiCliqueEdges}(b) illustrates the correlation between the overlapping level and link weight (including $w^N$ and $w^D$). An increase of the number of call $w^N_{ij}$ can be evidently seen with the growth of $w_{ij}^o$, which means that two mobile phone users belong to more cliques if they have more mutual calls. Such positive correlation is also observed between the overlapping level $w_{ij}^o$ and the call duration $w^D_{ij}$. The average sensitivity of call frequency on the overlapping level is higher for the SVDCN than for the DCN.

\subsection{Clique graph}

The clique graph $C(G)$ of an undirected graph $G$ is the intersection graph of the set of all cliques (i.e. the maximal complete subgraphs) of $G$ \cite{Prisner-1995}, where each clique is a node with the strength being its size and each edge has a weight being the size of the intersection of two cliques. Hence, the clique graph presents the interaction among different cliques, which can be viewed as ``friend circles''. The node strength stands for the size of the clique and the edge weight is the number of members shared by the two intersecting cliques.

The clique graph of the DCN has 1,162,832 nodes and 467,574,317 links, which means that each clique intersects with 804.2 cliques on average. The maximum degree is 17438 with the corresponding clique size or node strength being 21. The clique graph also has 2.97\% isolate cliques. For the SVDCN, the clique graph has 270,534 nodes and 14,099,314 links, indicating that each clique intersects with 104.2 cliques on average. The maximum degree is 2013 with the corresponding clique size or node strength being 6. The clique graph also has 5.16\% isolate cliques.

\begin{figure}[htp]
  \centering
  \includegraphics[width=0.99\textwidth]{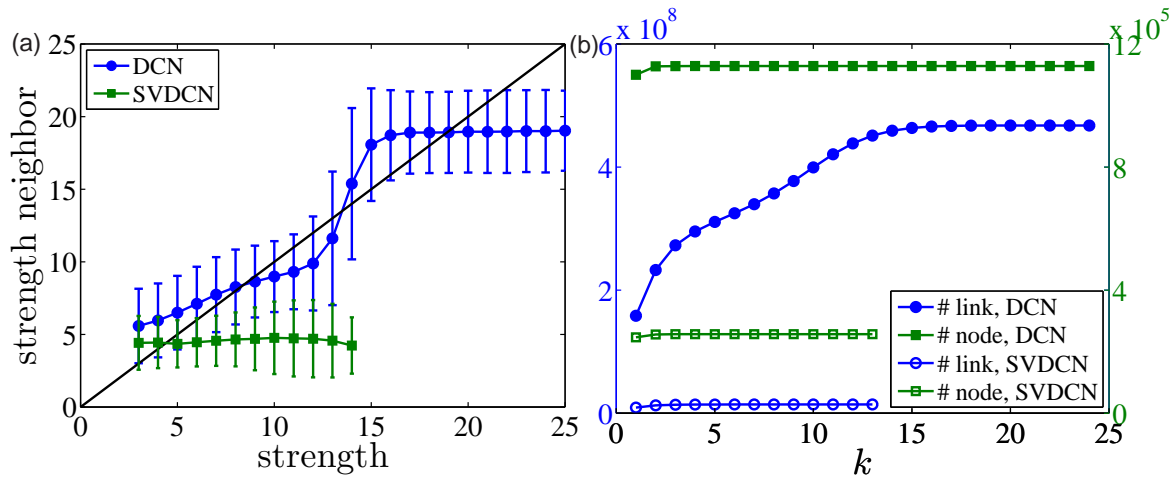}
  \caption{{\textbf{Properties of clique graphs of the two calling networks.}} (a) Average strength of the nearest neighbor nodes, $\langle{k_{\rm{nn}}}\rangle$, as a function of node strength $k$. The error bars stand for the standard deviations. (b) Giant component sizes (number of nodes and number of edges) of the $k$-overlap clique graph as a function of $k$.}
\label{Fig:CliqueGraph}
\end{figure}

Figure \ref{Fig:CliqueGraph}(a) shows the dependence of average node strength $\langle{k_{\rm{nn}}}\rangle$ of the nearest neighbors of a node on the node strength $k$ for the two calling networks. We find that, for the DCN, $\langle{k_{\rm{nn}}}\rangle \propto k/2$ when $k\leq12$, and $\langle{k_{\rm{nn}}}\rangle \propto 2.2k$ when $12\leq k\leq15$. It suggests that large (resp. small) cliques tend to interact with large (resp. small) cliques. When $k\geq16$, $\langle{k_{\rm{nn}}}\rangle$ is independent of $k$. For the SVDCN, $\langle{k_{\rm{nn}}}\rangle$ is almost independent of $k$.

For an integer $k \geq 1$, the $k$-overlap clique graph $C_k(G)$ of $G$ has all cliques of $G$ as its nodes, and two such cliques are adjacent whenever their intersection contains at most $k$ nodes \cite{Prisner-1995}. Figure \ref{Fig:CliqueGraph}(b) illustrates the dependence of the giant component size as a function of $k$. We find that, for both networks, the number of nodes of the giant component almost does not depend on the overlap $k$. The number of links of the giant component for the DCN increases with $k$, while that for the SVDCN is independent of the overlap $k$.

\section{Behaviors of mobile phone users}
\label{S1:Behavior}

\subsection{Distance among intra-clique users}

\begin{figure}[htp]
  \centering
  \includegraphics[width=0.99\textwidth]{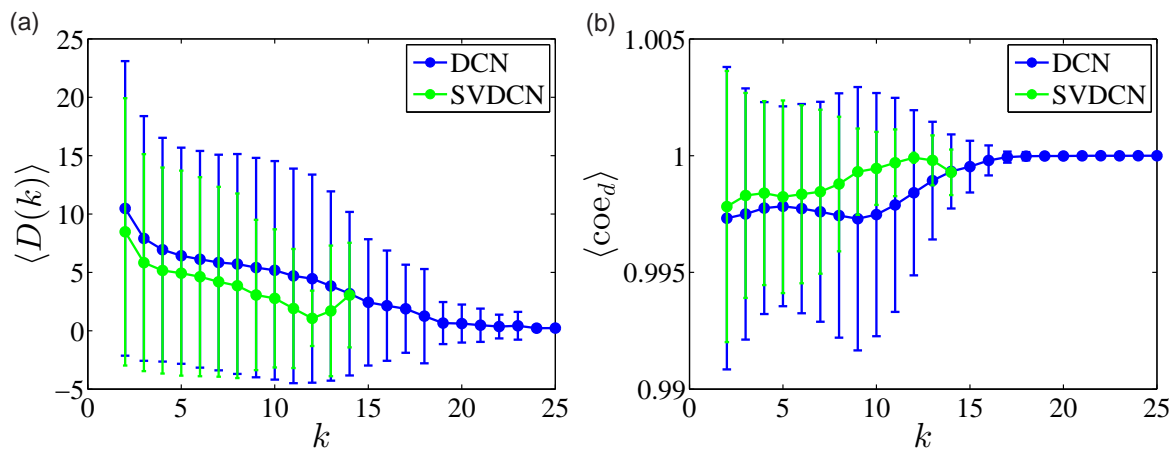}
  \caption{{\textbf{Distance among intra-clique mobile phone users for the DCN and the SVDCN.}} (a) The average inter-member distance $\langle{D(k)}\rangle$ is a decreasing function of clique size $k$. (b) The average center-to-Earth distance ${\rm{coe}}_d$ is an increasing function of clique size $k$}
  \label{Fig:Distance}
\end{figure}

We now investigate the distance among mobile phone users in the same clique. The map of Shanghai is divided into squares of length $0.0045^\circ$ (nearly 0.5 km). The most frequently visited (MFV) location (or square) of a user is identified by counting the number of days she visited the location. In each day, a user may visit several different locations. The Euclidean distance (in units of kilometer) between the most frequently visited locations of users $i$ and $j$ is calculated and we average these distances over all users in the same clique to obtain average inter-member distance $D(k)$ of a $k$-clique. The dependence of the average inter-member distance $\langle{D(k)}\rangle$ over all $k$-cliques against clique size $k$ is plotted in Fig.~\ref{Fig:Distance}  (a) for the two calling networks. The results for the two networks are basically the same except that the average distance among intra-clique users is slightly shorter for the SVDCN. The average distance decreases from about 10 km for $k=3$ to 0.5 km for $k=25$ and the standard deviation also decreases. In other words, the mobile phone users in the same clique of large size are geographically close to each other. For instance, if we plot the 25 users belonging to a 25-clique on the map, the 25 points cannot be distinguished from each other. We conjecture that these users belong to a same company at the Lujiazui Finance and Trade Zone (Pudong District, Shanghai).

%

The average inter-member distance is a measure reflecting geographical dispersion of members in a clique. One can also use the center-to-Earth distance to measure spatial dispersion of geolocated points on the Globe \cite{MartinBorregon-Aiello-Grabowicz-Jaimes-BaezaYates-2014-EPJDS}. As in Fig.~\ref{Fig:Distance} (b), one can observe that the center-to-Earth distance ${\rm{coe}}_d$ would be higher for large cliques than for small cliques, whereas the average distance is lower \cite{MartinBorregon-Aiello-Grabowicz-Jaimes-BaezaYates-2014-EPJDS}. Because all the mobile phone users are located in Shanghai, there is no doubt that ${\rm{coe}}_d$ for each clique is close to 1.

\subsection{Distribution of intercall durations}

The distribution of intercall durations of a mobile phone user is an important trait, which can be used to classify mobile phone users into several groups \cite{Jiang-Xie-Li-Podobnik-Zhou-Stanley-2013-PNAS}. Here we focus on the calls made among the mobile phone users in individual cliques. Specifically, we determine all the intraday intercall durations among the $k$ mobile phone users of a $k$-clique and obtain its empirical distribution.

\begin{figure}[htp]
  \centering
  \includegraphics[width=0.99\textwidth]{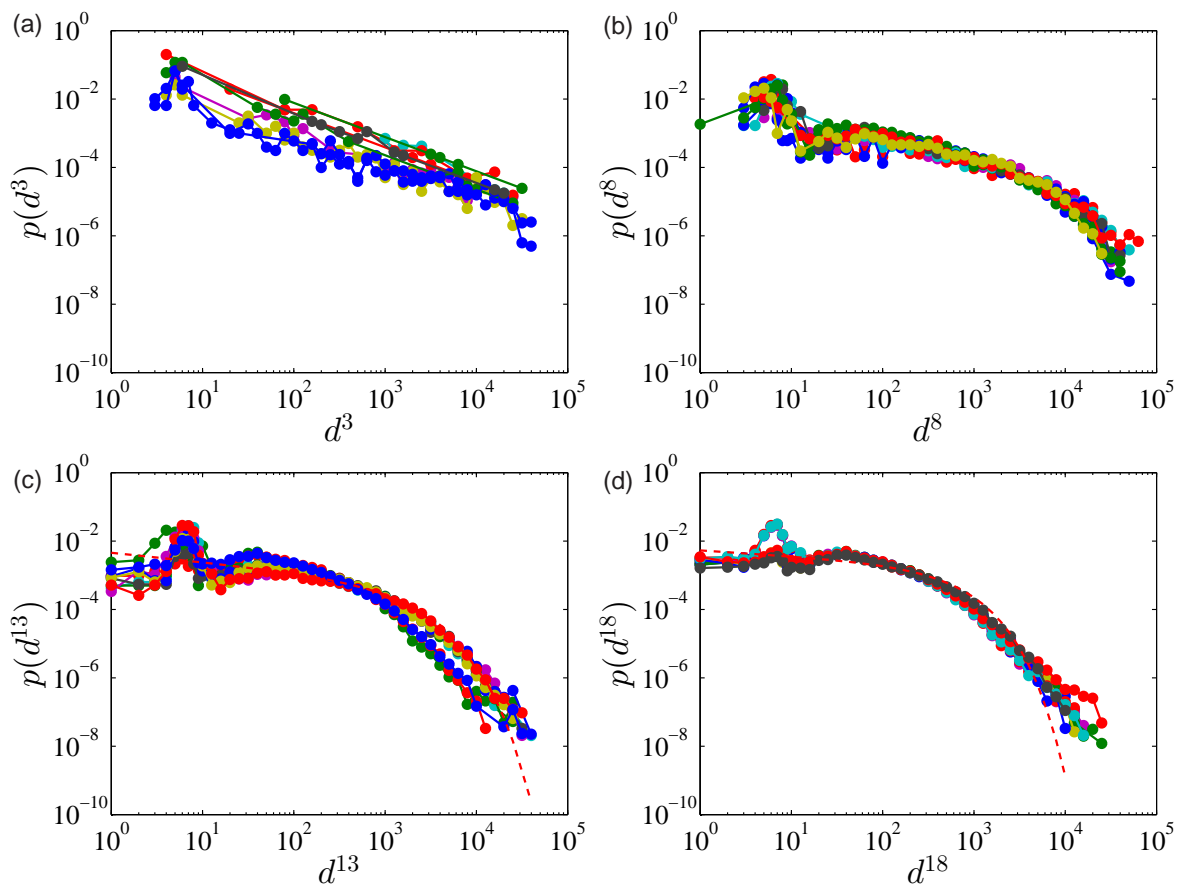}
  \caption{{\textbf{Distribution of intercall durations of all users in individual cliques.}} Each plot presents the probability distributions of intraday intercall durations of 10 randomly chosen cliques with different sizes: (a) $k=3$, (b) $k=8$, (c) $k=13$, and (d) $k=18$. Red dash lines in (c) and (d) refer to the fitted distribution for one of the cliques obtained by MLE method. The corresponding Weibull scale parameters are 0.73 and 0.83.}
  \label{Fig:PDF:ICDuration}
\end{figure}

Figure \ref{Fig:PDF:ICDuration} shows the probability distributions of the intraday intercall durations of 10 randomly chosen cliques for $k= 3$, 8, 13 and 18, respectively. For small-size cliques ($k=3$ in Fig. \ref{Fig:PDF:ICDuration}(a)), the intercall duration distributions have a power-law form. For relatively-mediate-size cliques ($k=8$ in Fig.~\ref{Fig:PDF:ICDuration}(b)), the distributions have a power law form with a fast cutoff. For mediate- and large-size cliques ($k=13$ and $k=18$ in Fig. \ref{Fig:PDF:ICDuration}(c) and (d)), the distributions are close to the Weibull distributions (red dash lines in figure). When the size of a clique is not small, its members are not likely to be abnormal users. Hence, the inter-call durations of these individual users are Weibull, and the mixture of these Weibull distributions are also Weibull \cite{Jiang-Xie-Li-Podobnik-Zhou-Stanley-2013-PNAS}.

\subsection{Classification of active cell phone users}

Classification of mobile phone users has important academic and commercial implications. Different approaches can be applied to classify users for different purposes \cite{Jiang-Xie-Li-Podobnik-Zhou-Stanley-2013-PNAS}, which usually results in different groups of users. Here we apply the k-means approach based on the $m$-clique degrees of each user. For each node $i$, we can obtain a vector of clique degrees ${\mathbf{X}}_i = \{k_i^{(2)}, k_i^{(3)}, ... ,k_i^{(25)}\}$, which defines the trait dimensions of mobile phone users. The distance between user $i$ and user $j$ is defined by the squared Euclidean distance $\parallel{\mathbf{X}}_i-{\mathbf{X}}_j\parallel$. In Ref.~\cite{Jiang-Xie-Li-Podobnik-Zhou-Stanley-2013-PNAS}, the 100,000 most active users are investigated. About 3.46\% of the users have a power-law distribution in the intraday intercall durations and about 73.34\% users have a Weibull distribution. Together with the calling diversity and out-degree of users, the power-law family can further be classified into three groups. In this work, we analyze these 76.8\% users. The DCN contains all these users, while the SVDCN has about 75000 users. We find that the users in the DCN can be classified into 6 groups and those in the SVDCN can be categorized into four groups, where the number of groups is determined by the silhouette value. There is no direct connection between these two kinds of groups, which neither do not have direct relationship with the four groups identified in Ref.~\cite{Jiang-Xie-Li-Podobnik-Zhou-Stanley-2013-PNAS}.

\begin{figure}[htp]
  \centering
  \includegraphics[width=0.99\textwidth]{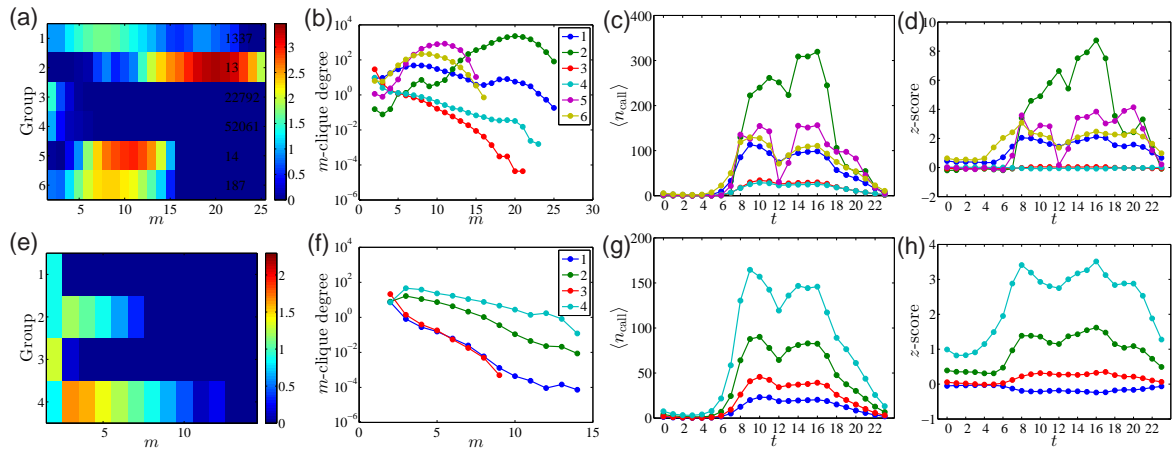}
  \caption{{\textbf{Classification of active mobile phone uses by using the k-means approach.}} The first row is for the DCN and the second row is for the SVDCN. (a, e) Categorical attribute for the center of each group. The color bars show the 10-based logarithms of the $m$-clique degrees. (b, f) Clique degree profile as a function of clique size for each identified group. (c, g) Intraday pattern in the average number of calls $\langle{n_{\mathrm{call}}}\rangle$ for the users in each identified group. (d, h) Corresponding $z$-score for the average number of calls in (c) and (g).
}
\label{Fig:CliqueDegree:kmeans}
\end{figure}

The categorical attribute for the center of each group is shown in Fig. \ref{Fig:CliqueDegree:kmeans} (a) and (e). One can list descending orders from the highest clique degree for the group centers, which are 2, 5, 6, 1, 3, 4 for DCN and 4, 2, 3, 1 for SVDCN. They are the same as the orders shown in table \ref{Table:GroupProperty}. We also show the profile of $m$-clique degrees for each group in Fig. \ref{Fig:CliqueDegree:kmeans}(b) and (f). It is found that the profiles of the DCN are more complicated than those of the SVDCN. For the SVDCN, the $m$-clique degree decreases exponentially with $m$ for each group. Group 3 seems similar to group 1 except that the users in group 3 do not have cliques with $m>9$. The main difference between these groups is the exponential decay rate. For the DCN,the profiles have very different shapes. Two (group 3 and group 4) of them decrease monotonically, three profiles (groups 2, 5 and 6) have a bell shape, and the last one (group 1) is bimodal. As shown in Fig. \ref{Fig:CliqueDegree:kmeans}(c) and (g), the members in different groups had different average intraday numbers of initiated calls. Users with high $m$-clique degrees are found to have initiated more calls and have large $z$-scores when compared with all users as shown in Fig. \ref{Fig:CliqueDegree:kmeans}(d) and (h).

Table \ref{Table:GroupProperty} presents several other properties of the mobile phone users in different groups. For the DCN, all the quantities for group 3 and group 4 are comparable to each other and significantly small than other groups. Interestingly, the 6 groups have the same order in these quantities except for the in-degree and the out-degree. For example, users in group 1 have the largest in-coming and out-going call strength, coreness and clustering coefficient, which are the most active users. For the SVDCN, the 4 groups rank in a perfect order for all the quantities. We find that group 4 and group 2 have comparable quantities, while group 3 and group 1 also have comparable quantities. However, the discrepancy between group 2 and group 3 is remarkably evident. These observations are consistent with the results in Fig. \ref{Fig:CliqueDegree:kmeans}. One can observed that people with higher clique degrees (group 2 in Fig. \ref{Fig:CliqueDegree:kmeans}) will have strong social desires. For example, people in group 2 make more calls $s^N$ and tend to be the center of group (higher coreness $cn$).

\begin{table}[htp]
  \centering
  \caption{{\textbf{Node properties for the 6 groups in the DCN and the 4 groups in the SVDCN.}} $k_{\rm{out}}$ is the average out-degree and $k_{\rm{in}}$ is the average in-degree. $s^N_{\rm{out}}$ and $s^N_{\rm{in}}$ are the number-weighted strengthes of the out-links and in-links. $cn_{\rm{out}}$ and $cn_{\rm{in}}$ are the average node coreness of out-links and in-links. $cc_{\rm{dir}}$ and $cc_{\rm{undir}}$ are the average node clustering coefficients for directed links and non-directed links.}
  \label{Table:GroupProperty}
  \begin{tabular}{r*{8}{r}}
  \hline
  \hline
  group & $k_{\rm{out}}$ & $k_{\rm{in}}$ & $s^N_{\rm{out}}$ & $s^N_{\rm{in}}$ & $cn_{\rm{out}}$ & $cn_{\rm{in}}$ & $cc_{\rm{dir}}$ & $cc_{\rm{undir}}$\\
  \hline
Panel A: DCN \\
  \hline
  2 & 82.31 & 87.54 & 2785.15 & 2499.23 & 52 & 49 & 0.69 & 0.73\\
  5 & 97.43 & 98.86 & 1584.93 & 1404.57 & 39 & 41 & 0.47 & 0.52\\
  6 & 108.43 & 113.86 & 1405.06 & 1316.06 & 30.41 & 30.55 & 0.32 & 0.34\\
  1 & 88.19 & 90.65 & 1172.48 & 1015.7 & 23.89 & 23.73 & 0.26 & 0.27\\
  3 & 40.42 & 40.02 & 342.16 & 324.37 & 8.83 & 8.66 & 0.03 & 0.03\\
  4 & 22.33 & 19.79 & 308.5 & 262.03 & 7.3 & 6.96 & 0.1 & 0.09\\
  \hline
Panel B: SVDCN \\
  \hline
  4 & 40.65 & 40.02 & 1734.79 & 1453.54 & 12.68 & 12.88 & 0.37 & 0.4\\
  2 & 25.2 & 25 & 904.64 & 814.47 & 9.04 & 9.12 & 0.35 & 0.35\\
  3 & 17.39 & 17.57 & 423.43 & 386.34 & 4.38 & 4.33 & 0.07 & 0.07\\
  1 & 7.76 & 7.44 & 224.38 & 191.18 & 3.06 & 2.92 & 0.11 & 0.1\\
  \hline
  \hline
  \end{tabular}
\end{table}

\section{Discussion and conclusion}
\label{S1:Discussion}

In summary, we have performed a comparative investigation of the structural properties of cliques in the original directed calling networks and its statistically validated Bonferroni network. The statistical properties of the cliques of the two calling networks are qualitatively similar but also exhibit idiosyncratic behaviors quantitatively. We found that the members in large cliques are spatially close to each other, indicating the importance of geographic information on the local structure of social networks. Based on the clique degree profile of each mobile phone user, the most active users in the two calling networks can be classified in to several groups with different calling behaviors.

The structural properties of calling network cliques reported in this work are relevant to the social interactions among humans, which may contain important information and deserves further investigations. For instance, the properties of average inter-member distance can be partially explained by the social brain hypothesis \cite{Dunbar-1998-EA}, besides the conjecture of belonging to a same organization such as a company for all clique members. Because human' processing power limits social group size \cite{DavidBarrett-Dunbar-2013-PRSB} and to maintain social ties is also costly in resources such as time and money, it is rational to observe shorter clique diameters for large cliques as shown in Fig.~\ref{Fig:Distance}.

Cliques of size larger than four are a kind of meso-structure larger than motifs and smaller than modular communities. It is well documented that motifs and communities have important effects on the behaviors and functions of the associated complex systems. We thus argue that our empirical findings on the statistical properties of cliques in human communication networks may play important roles in human behaviors and affect social functions. For instance, social network is an important factor influencing the evolution of cooperation \cite{Nowak-2005-Science,Perc-Szolnoki-2010-Bs,Perc-GomezGardenes-Szolnoki-Floria-Moreno-2013-JRSI,Szolnoki-Perc-2013-PRX}. It can be conjectured that these locally dense structures will be also very relevant to human's cooperative behavior. Studies on such kind of topics have essential scientific significance, which is however beyond the scope of this work.

\section*{Acknowledgments}

This work was partially supported by the National Natural Science Foundation of China (11205057), the Ph.D. Programs Foundation of Ministry of Education of China (20120074120028), the Fok Ying Tong Education Foundation (132013), and the Humanities and Social Sciences Fund of the Ministry of Education of China (09YJCZH040).

%

\providecommand{\newblock}{}

\end{document}